\documentclass[prl,twocolumn,twoside,preprintnumbers,superscriptaddress]{revtex4-1}
\usepackage{physics}
\usepackage{amsmath}
\usepackage{amssymb}
\usepackage{hyperref}
\usepackage{xspace}
\usepackage{slashed}
\hypersetup{
    colorlinks=true,       
    linkcolor=blue,          
    citecolor=blue,        
    filecolor=blue,      
    urlcolor=blue           
}
\usepackage{xcolor}
\usepackage{color}
\usepackage{graphicx}  %
\usepackage{bm}  %
\usepackage{amsmath}   %
\usepackage{graphics}
\usepackage{color}
\usepackage{colortbl}
\usepackage{ulem}

\newcommand{\dslash}[1]{#1 \llap{/\kern-0.5pt}}
\newcommand{\Dslash}[1]{#1 \llap{/\kern+1.5pt}}
\newcommand{\DDslash}[1]{#1 \llap{/\kern+2.3pt}}
\newcommand{\dslashh}[1]{#1 \llap{/\kern+1pt}}

\newcommand{\bea}{\begin{eqnarray}}
\newcommand{\eea}{\end{eqnarray}}
\newcommand{\be}{\begin{equation}}
\newcommand{\ee}{\end{equation}}
\newcommand{\bma}{\begin{pmatrix}}
\newcommand{\ema}{\end{pmatrix}}

\newcommand{ \mysmall}[1]{\scriptscriptstyle #1} 

\allowdisplaybreaks[1]

\begin{document}

\title{Probing the muon {\boldmath $g$-2} anomaly with the Higgs boson at a Muon Collider}

\author{Dario Buttazzo}
\affiliation{Istituto Nazionale di Fisica Nucleare, Sezione di Pisa, I-56127 Pisa, Italy}
\author{Paride Paradisi}
\affiliation{Dipartimento di Fisica e Astronomia `G. Galilei', Universit\`a di Padova, Italy}
\affiliation{Istituto Nazionale Fisica Nucleare, Sezione di Padova, I--35131 Padova, Italy}

\begin{abstract}
We point out that heavy new physics contributions in leptonic dipole moments and high-energy cross-sections of lepton pairs into Higgs bosons and photons are 
connected model-independently. In particular, we demonstrate that a muon collider, running at center-of-mass energies of several TeV, can provide a unique test 
of new physics in the muon $g$-2 through the study of high-energy processes such as $\mu^+\mu^- \to h \gamma$.
This high-energy test would be of the utmost importance to shed light on the long-standing muon $g$-2 anomaly as it is not affected by the hadronic and 
experimental uncertainties entering the current low-energy determination of the muon $g$-2.  
Furthermore, we show that  the current bound on the muon electric dipole moment can be improved by three orders of magnitude, down to ${\rm few} \times 10^{-22}\,e\,$cm.
\end{abstract}

\maketitle

\section{I. Introduction}
The anomalous magnetic moment of the muon has provided, over the last ten years, an enduring hint for new physics (NP). 
The experimental value of $a_\mu \!=\! (g_\mu \!-\! 2)/2$ from the E821 experiment at BNL~\cite{Bennett:2006fi} was recently confirmed by the E989 
experiment at Fermilab~\cite{Abi:2021gix}, yielding the experimental average $a_\mu^{\mysmall \rm EXP} \!=\! 116592061(41) \!\times\! 10^{-11}$. 
The comparison of this value with the Standard Model (SM) prediction $a_\mu^{\mysmall \rm SM} \!=\! 116591810(43) \times 10^{-11}$~\cite{Aoyama:2020ynm} 
shows an interesting $4.2\,\sigma$ discrepancy~\cite{Abi:2021gix}
\be
\Delta a_\mu = a_\mu^{\mysmall \rm EXP}-a_\mu^{\mysmall \rm SM} = 251 \, (59) \times 10^{-11}\,.
\label{eq:gmu}
\ee
The forthcoming runs of the E989 experiment plan to reduce the experimental uncertainty by a factor of four. Moreover, a completely new low-energy 
approach to measuring the muon $g$-2 is being developed by the E34 collaboration at J-PARC~\cite{Abe:2019thb}. On the theory side, there is also 
an ongoing effort to reduce the leading SM uncertainty stemming from hadronic corrections~\cite{lattice}.  

Given the difficulty of controlling all these effects at the required level of precision, we think it is crucial to have an independent test of NP in the muon $g$-2, 
not affected by the hadronic and experimental uncertainties entering the current low-energy determination of the muon $g$-2.

Incidentally, the observed muon $g$-2 discrepancy can be accommodated by a NP effect of the same size as the 
SM weak contribution $\sim 5\,G_{\rm F} m_\mu^2 / 24 \sqrt{2}\pi^2 \approx 2 \times 10^{-9}$~\cite{Aoyama:2020ynm}.
Therefore, a very natural explanation of eq.~(\ref{eq:gmu}) could be achieved within weakly interacting NP scenarios 
emerging at a scale $\Lambda$ close to the electroweak scale.
Remarkably, this possibility could be connected with the solution of the {\it hierarchy problem}
and could provide, at the same time, a WIMP dark matter candidate.
Unfortunately, the lack for new particles at LEP and LHC strongly disfavours this interpretation.
As a result, two possibilities seem to emerge to solve the muon $g$-2 anomaly while avoiding the stringent LEP and LHC bounds. 
Either NP is very light ($\Lambda \lesssim 1~$GeV) and feebly coupled to SM particles, see e.g.~\cite{Marciano:2016yhf}, 
or NP is very heavy ($\Lambda \gg 1~$TeV) and strongly coupled. Here, we take the second direction.

Heavy NP contributions to the muon $g$-2 arise from the dimension-6 dipole operator 
$\left(\bar\mu_L \sigma_{\mu\nu} \mu_R\right) H F^{\mu\nu}$~\cite{Buchmuller:1985jz} 
where $H \!=\! v+h/\sqrt{2}$ contains both the Higgs boson field $h$ and its vacuum expectation value $v \!=\! 174~$GeV
and $F^{\mu\nu}$ is the electromagnetic field strenght tensor.
After electroweak symmetry breaking $H \!\to v$ and we obtain the prediction 
$\Delta a^{\rm\mysmall NP}_\mu \sim (g^2_{\mysmall\rm NP}/ 16\pi^2) \times (m_\mu v/\Lambda^2)$, 
where $g_{\rm\mysmall NP}$ is the typical coupling of the NP sector.
Therefore, the NP chiral enhancement $v/m_\mu \sim 10^3$ with respect to the SM weak contribution, together with the assumption 
of a new strong dynamics with $g_{\rm\mysmall NP} \sim 4 \pi$, bring the sensitivity of the muon $g$-2 to NP scales of order $\Lambda \sim 100\,$TeV~\cite{masses}.
 
Directly detecting new particles at such high scales is far beyond the capabilities of any foreseen collider. Moreover, even assuming the discovery of new particles 
by their direct production~\cite{Capdevilla:2020qel}, it would be very hard, if not impossible, to unambiguously associate them to $\Delta a_\mu$. In other words, 
it would be desirable to test the muon $g$-2 anomaly model-independently.

In this work, we argue that a muon collider (MC) running at energies $E$ of several TeV would represent the only machine enabling to probe NP in the muon $g$-2 in a 
completely model-independent way. Indeed, the very same dipole operator that generates $\Delta a_\mu$ unavoidably induces also a NP 
contribution to the scattering process $\mu^+\mu^- \to h \gamma$. Measuring the cross-section for this process would thus be equivalent to measuring $\Delta a_\mu$. 
This would however be a direct determination of the NP contribution, not hampered by the hadronic uncertainties that affect the SM prediction of $a_\mu$.

At first sight, it could seem impossible to be sensitive to such a tiny value of $\Delta a_\mu \sim 10^{-9}$ at a collider experiment. 
However, analogously to the case of weak interaction cross-sections in the effective Fermi theory, the cross-section for $\mu^+\mu^-\to h\gamma$ as induced by the 
effective dipole operator grows with the square of the collider energy. As a result, a high-energy measurement with $\mathcal{O}(1)$ precision will be sufficient to disentangle 
NP effects from the SM background.
This is the first example in high-energy particle physics of a sensitivity to a magnetic moment at this level, several orders of magnitude below all the other current and projected 
collider constraints. In order to reach such tiny values of $\Delta a_\mu$ it is however crucial to accelerate the muon pairs to the highest possible multi-TeV energies.

We stress that our results are valid for $E \ll \Lambda$ where the effective field theory (EFT) description is justified.

A high-energy MC with the luminosity needed for particle physics experiments~\cite{Delahaye:2019omf} is currently not feasible. Nevertheless, several efforts to overcome the technological challenges are ongoing~\cite{muoncollider}, and it is crucial to explore the broad physics potential of such a machine in order to pave the road for the forthcoming accelerator and detector studies. A MC is the ideal machine to search for NP at the highest possible energies, both directly and indirectly. Indeed, muons could in principle be accelerated to multi-TeV energies, as their larger mass greatly suppresses synchrotron radiation compared to the electron-positron case. Furthermore, the physics reach of the 
MC overtakes that of a proton-proton collider of the same energy since all of the beam energy is available for the hard collision, compared to the fraction of the proton energy 
carried by the partons: a MC in the 10 TeV range has roughly the same energy available for hard scatterings as a 100 TeV hadron collider~\cite{Delahaye:2019omf}.  

The physics case of a high-energy determination of $\Delta a_\mu$, which is unique of a MC, represents a striking example of the complementarity and interplay of the high-energy 
and high-intensity frontiers of particle physics. At the same time, it highlights the far reaching potential of a MC, that offers a new powerful way to probe NP which is complementary 
both to direct searches for new particles, and to the indirect tests conducted at low energy through high-precision experiments.
 
The paper is organised as follows. In section II, we introduce the SM effective field theory (SMEFT), containing operators up to dimension-6, 
contributing to $a_\ell$. After performing a one-loop calculation of $a_\ell$ in such EFT, in section~III, we study the high-energy processes 
at a MC which are sensitive to the same NP effects entering $a_\ell$. In section~IV, we comment on the possibility of measuring the rare 
Higgs decays $h \to \ell^+\ell^-\gamma$ (with $\ell = \mu,\tau$) that are induced by the same dipole operator generating $a_\ell$. 
The huge number of Higgs bosons that could be produced at a MC~\cite{Costantini:2020stv} could in principle allow the measurement of 
these rare processes, and thus the extraction of $a_\ell$.

\section{II. The muon $g$-2 in the SMEFT}
%
\begin{figure}
\centering%
\includegraphics[width=0.45\textwidth]{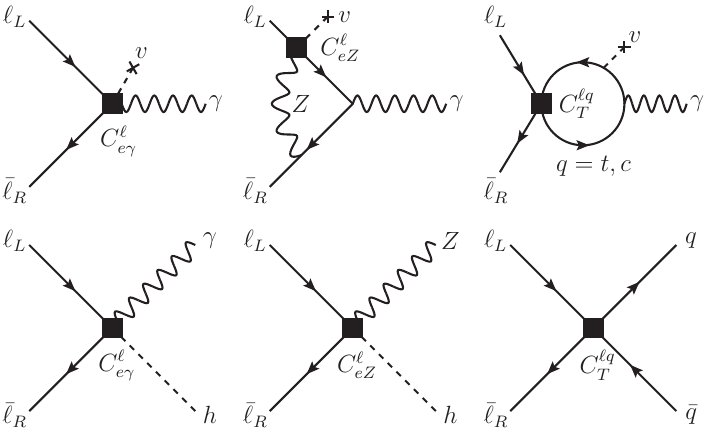}
\caption{{\it Upper row:} Feynman diagrams contributing to the leptonic $g$-2 up to one-loop order in the Standard Model EFT.
{\it Lower row:} Feynman diagrams of the corresponding high-energy scattering processes.
Dimension-6 effective interaction vertices are denoted by a square.}
\label{fig:feyn}
\end{figure}
%
New interactions emerging at a scale $\Lambda$ larger than the electroweak scale can be described at energies $E \ll \Lambda$ 
by an effective Lagrangian containing non-renormalizable $SU(3)_c \otimes SU(2)_L \otimes U(1)_Y$ invariant operators.
Focusing on the leptonic $g$-2, the relevant effective Lagrangian contributing to them, up to one-loop order, reads~\cite{Buchmuller:1985jz}
\begin{align}
\mathcal{L} &= 
\frac{C^\ell_{eB}}{\Lambda^2}
\left( \bar\ell_L \sigma^{\mu\nu}e_{R}\right) \! H B_{\mu\nu} + 
\frac{C^\ell_{eW}}{\Lambda^2}
\left( \bar\ell_L \sigma^{\mu\nu} e_{R} \right) \! \tau^I \! H W_{\mu\nu}^I 
\nonumber\\
&+ \frac{C^\ell_{T}}{\Lambda^2}( \overline{\ell}^a_L\sigma_{\mu\nu}e_{R}) \varepsilon_{ab} (\overline{Q}^b_L\sigma^{\mu\nu} u_{R}) 
+ h.c.
\label{eq:L_SMEFT}
\end{align} 
where it is assumed that the NP scale $\Lambda \gtrsim 1$ TeV.
The Feynman diagrams relevant for the leptonic $g$-2 are displayed in figure~\ref{fig:feyn}. 
They lead to the following result
\begin{align}
\Delta a_\ell  &\simeq \frac{4m_\ell v}{e\Lambda^2} \, 
\bigg(
C^\ell_{e\gamma} - \frac{3\alpha}{2\pi} \frac{c^2_W \!-\! s^2_W}{s_W c_W} \,C^\ell_{eZ} \log\frac{\Lambda}{m_Z}
\bigg)
\nonumber\\
& - \sum_{q=c,t} \frac{4m_\ell m_q}{\pi^2} \frac{C_T^{\ell q}}{\Lambda^2}\,
\log\frac{\Lambda}{m_q},
\label{eq:Delta_a_ell}
\end{align}
where $s_W$, $c_W$ are the sine and cosine of the weak mixing angle, $C_{e\gamma}=c_W C_{eB} - s_W C_{eW}$ and
$C_{eZ} = -s_W C_{eB} - c_W C_{eW}$.
Additional loop contributions from the operators $H^\dag H W_{\mu\nu}^IW^{I\mu\nu}$, $H^\dag H B_{\mu\nu}B^{\mu\nu}$, 
and $H^\dag \tau^I H W_{\mu\nu}^I B^{\mu\nu}$ are suppressed by the lepton Yukawa couplings and can be neglected.
Moreover, in eq.~(\ref{eq:Delta_a_ell}), we assumed for simplicity that $C_{eB}$, $C_{eW}$ and $C_{T}$ are real. 
Since only the first two operators of eq.~(\ref{eq:L_SMEFT}) generate electromagnetic dipoles at tree-level, 
we include their one-loop renormalization effects to $C^\ell_{e\gamma}$ 
\begin{align}
\!\!\! C^\ell_{e\gamma}(m_\ell) \simeq C^\ell_{e\gamma}\!(\Lambda)\left(1 \!-\! \frac{3y^2_t}{16\pi^2} \log\frac{\Lambda}{m_t}
\!-\! \frac{4 \alpha}{\pi} \log\frac{m_t}{m_\ell}\right).
\label{eq:running_Cegamma}
\end{align}
In order to see where we stand, let us determine the NP scale probed by $\Delta a_\ell$.
From eq.~(\ref{eq:Delta_a_ell}) we find that
\begin{align}
\!\!
\frac{\Delta a_\mu}{3 \!\times\! 10^{-9}} \!\approx \! \left( \frac{250 \, {\rm TeV}}{\Lambda} \right)^{\!\!2} \!\!\!
\left(C^\mu_{e\gamma} \!-\! 0.2 C^{\mu t}_T \!\!-\! 0.001 C^{\mu c}_T \!\!-\! 0.05 C^{\mu}_{eZ}\right). 
\nonumber
\end{align}

\medskip

A few comments are in order: 
\begin{itemize}
\item The $\Delta a_\mu$ discrepancy can be solved for a NP scale up to $\Lambda\approx 250~$TeV.
This requires a strongly coupled NP sector where $C^\mu_{e\gamma}$ and/or $C^{\mu t}_T \sim g^2_{\rm\mysmall NP}/16\pi^2 \sim 1$ 
and a chiral enhancement $v/m_\mu$ compared with the weak SM contribution~\cite{gm2_e}.
For such large values of $\Lambda$ direct NP particle production is beyond the reach of any foreseen collider. However, as we shall see, 
the physics responsible for $\Delta a_\mu$ can still be tested through high-energy processes such as $\mu^+\!\mu^- \!\to h\gamma$ or 
$\mu^+\!\mu^- \!\to q\bar{q}$ (with $q=c,t$).
\item If the underlying NP sector is weakly coupled, $g_{\rm\mysmall NP}\lesssim 1$, then
$C^\mu_{e\gamma}$ and $C^{\mu t}_T \lesssim 1/16\pi^2$, implying $\Lambda\lesssim 20~$TeV to solve the $\Delta a_\mu$ anomaly.
In this case, a MC could still be able to directly produce NP particles~\cite{Capdevilla:2020qel}. 
Yet, the study of the processes $\mu^+\mu^-\to h\gamma$
and $\mu^+\mu^-\to q\bar{q}$ could be crucial to reconstruct the effective dipole vertex $\mu^+\mu^-\gamma$.
\item If the NP sector is weakly coupled, and further $\Delta a_\mu$ scales with lepton masses as the SM weak contribution,
then $\Delta a_\mu \sim m^2_\mu/16\pi^2\Lambda^2$.
Here, the experimental value of $\Delta a_\mu$ can be accommodated only provided that $\Lambda \lesssim 1~$TeV.
For such a low NP scale the EFT description breaks down at the typical multi-TeV MC energies, 
and new resonances cannot escape from direct production.
\end{itemize}

\section{III. High-energy probes of the muon $g$-2} 
The main contribution to $\Delta a_\mu$ comes from the dipole operator $O_{e\gamma}=\left(\bar\ell_L \sigma_{\mu\nu} e_R\right) H F^{\mu\nu}$ 
when after electroweak symmetry breaking $H\to v$. The same operator also induces a contribution to the process $\mu^+\mu^- \to h \gamma$ 
that grows with energy (see figure~\ref{fig:feyn}), and thus can become dominant over the SM cross-section at a very high-energy collider. 
Assuming that $m_h \ll \sqrt{s}$, which is an excellent approximation at a MC, we find the following differential cross-section
\begin{equation}\label{eq:sigma_hV}
\frac{d\sigma_{h\gamma}}{d\cos\theta} = 
\frac{|C^\mu_{e\gamma}|^2}{\Lambda^4}\frac{s}{64\pi}\left(1-\cos^2\theta \,\right)
\end{equation}
where $\cos\theta$ is the photon scattering angle. 
Notice that there is an identical contribution also to the process $\mu^+\mu^- \!\to\! Z\gamma$ since $H$ contains the longitudinal polarizations of the $Z$.
The total $\mu^+\mu^- \!\to  h\gamma$ cross-section is
\begin{align}
\sigma_{h\gamma} \!=\! \frac{s}{48\pi}\frac{|C^{\mu}_{e\gamma}|^2}{\Lambda^4}\!
\approx  0.7\, {\rm ab} \left(\frac{\sqrt{s}}{30\, {\rm TeV}}\right)^{\!2} \!\! \left(\frac{\Delta a_\mu}{3 \times 10^{-9}} \right)^{\!2}
\end{align}
where in the last equation we assumed no contribution to $\Delta a_\mu$ other than the one from $C^{\mu}_{e\gamma}$. 
Moreover, we included running effects for $C^{\mu}_{e\gamma}$, see eq.~(\ref{eq:running_Cegamma}), from a scale 
$\Lambda \approx 100$~TeV. Given the scaling with energy of the reference integrated luminosity~\cite{Delahaye:2019omf}
\begin{align}
\mathcal{L} = \left(\frac{\sqrt{s}}{10 \,{\rm TeV}} \right)^{2} \! \times 10 \,{\rm ab}^{-1}
\label{eq:luminosity_MC}
\end{align}
one gets about 60 total $h\gamma$ events at $\sqrt{s}=30~$TeV.

\begin{figure}
\centering%
\includegraphics[width=0.48\textwidth]{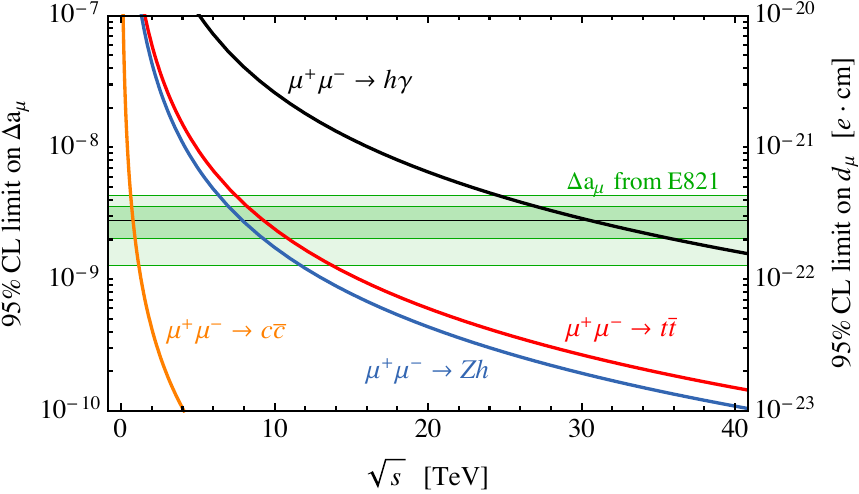}
\caption{95\% C.L.\ reach on the muon anomalous magnetic moment $\Delta a_\mu$, as well as on the muon EDM $d_\mu$, 
as a function of the collider center-of-mass energy $\sqrt{s}$, from the processes $\mu^+\mu^-\to h\gamma$ (black), $\mu^+\mu^-\to hZ$ (blue), 
$\mu^+\mu^-\to t \bar t$ (red), and $\mu^+\mu^-\to c \bar c$ (orange).}
\label{fig:hgamma}
\end{figure}

The SM irreducible $\mu^+\mu^- \to h\gamma$ background is small. 
The dominant contribution arises at one-loop~\cite{Abbasabadi:1995rc} due to the muon Yukawa coupling suppression of the tree-level part,
%
$\sigma_{h\gamma}^{\rm SM} \approx 2 \times 10^{-2}\,{\rm ab}\,\big(\frac{30\,{\rm TeV}}{\sqrt{s}}\big)^{\!2}$, 
%
and can be neglected for $\sqrt{s} \gg$~TeV.
The main source of background comes from $Z\gamma$ events, where the $Z$ boson is incorrectly reconstructed as a Higgs. 
This cross-section is large, due to the contribution from transverse polarizations,
\begin{align}
\frac{d\sigma_{Z\gamma}}{d\cos\theta} = \frac{\pi\alpha^2}{4s}\frac{1+\cos^2\theta}{\sin^2\theta}\frac{1-4 s_W^2 + 8 s_W^4}{s_W^2 c_W^2}.
\end{align}
There are two ways to isolate the $h\gamma$ signal from the background: by means of the different angular distributions of the two processes -- the SM $Z\gamma$ peaks in 
the forward region, while the signal is central -- and by accurately distinguishing $h$ and $Z$ bosons from their decay products, e.g. by precisely reconstructing their invariant mass.

To estimate the reach on $\Delta a_\mu$ we consider a cut-and-count experiment in the $b\bar b$ final state, which has the highest signal yield (with branching ratios $\mathcal{B}(h\to b\bar b)~=~0.58$, $\mathcal{B}(Z\to b\bar{b})~=~0.15$). The significance of the signal -- defined as $N_S/\sqrt{N_B+N_S}$, with $N_{S,B}$ the number of signal and background events -- is maximized in the central region 
$|\!\cos\theta| \lesssim 0.6$. At 30 TeV one gets
\begin{align}
\sigma_{h\gamma}^{\rm cut} &\approx 0.53\, {\rm ab} \,\bigg(\frac{\Delta a_\mu}{3 \times 10^{-9}} \bigg)^{\!2}, & ~~\sigma_{Z\gamma}^{\rm cut} &\approx 82\,{\rm ab}\,.
\end{align}
Requiring at least one jet to be tagged as a $b$, and assuming a $b$-tagging efficiency $\epsilon_b = 80\%$, we find that a value $\Delta a_\mu = 3\!\times\! 10^{-9}$ can be tested at 95\% C.L.\ at a 30 TeV collider if the probability of reconstructing a $Z$ boson as a Higgs is less than 10\%. The resulting number of signal events is $N_S = 22$, and $N_S/N_B = 0.25$.
In figure~\ref{fig:hgamma} we show as a black line the 95\% C.L.\ reach from $\mu^+\mu^-\to h\gamma$ 
on the anomalous magnetic moment as a function of the collider energy. Note that since the number of signal events scales as the fourth power of the center-of-mass energy, only a collider with $\sqrt{s} \gtrsim 30$~TeV will have the sensitivity to test the $g$-2 anomaly.
%

The $Z$-dipole operator $O_{eZ}=\left(\bar\ell_L \sigma_{\mu\nu} e_R\right) H Z^{\mu\nu}$ contributes to $\Delta a_\mu$ at one loop, 
and generates also the process $\mu^+\mu^- \to Zh$ (see figure~\ref{fig:feyn}) with the same cross-section of eq.~\eqref{eq:sigma_hV} 
with $\gamma\leftrightarrow Z$, so that
\begin{equation}
\sigma_{Zh} \approx 38 \, {\rm ab} \, \left(\frac{\sqrt{s}}{10 \, {\rm TeV}}\right)^{\!2} \!\! \left(\frac{\Delta a_\mu}{3 \times 10^{-9}} \right)^{\!2}.
\label{eq:sigma_hZ}
\end{equation}
Here we assume that only $O_{eZ}$ contributes to $\Delta a_\mu$: it should be stressed that this corresponds 
to an unnatural scenario, where the coefficients $C_{eB}$ and $C_{eW}$ conspire to cancel out the tree-level contribution from $O_{e\gamma}$. 
It is nevertheless meaningful to derive the constraint from high-energy scattering on the $Z$-dipole contribution to the 
$g$-2. The cross-section in eq.~\eqref{eq:sigma_hZ} has to be compared to the SM irreducible background given by
%
$\sigma_{Zh}^{\rm SM} \approx 122 \,{\rm ab}\,\big(\frac{10\,{\rm TeV}}{\sqrt{s}}\big)^{\!2}$.
%
Considering again the $h\to b\bar{b}$ channel, together with hadronic decays of the $Z$, one gets the 95\% C.L.\ limit shown in figure~\ref{fig:hgamma} as a blue line.

\begin{figure}
\centering%
\includegraphics[width=0.45\textwidth]{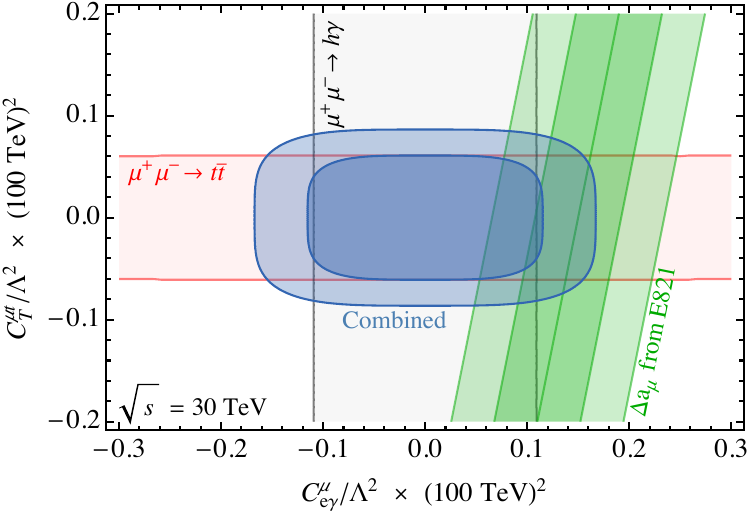}
\caption{
Constraints on the Wilson coefficients $C_{e\gamma}^\mu$ and $C_T^{\mu t}$ from $\mu^+\mu^- \to h\gamma$ and $\mu^+\mu^- \to t\bar{t}$ at a muon collider with $\sqrt{s} = 30$~TeV. The shaded regions are 68\% and 95\% C.L. contours, the individual 1$\sigma$ limits are also shown.
}
\label{fig:2D}
\end{figure}

Next, we derive the constraints on the semi-leptonic operators. The operator $O_T^{\mu t}$ that enters 
$\Delta a_\mu$ at one loop can be probed by $\mu^+\mu^-\to t\bar t$ (see figure~\ref{fig:feyn}). Its contribution 
to the cross-section is
\begin{align}
\label{eq:mumu_tt}
\!\!\!\sigma_{t\bar{t}} =\! \frac{s}{6\pi} \frac{|C^{\mu t}_{T}|^2}{\Lambda^4} N_c 
\approx 
58 \, {\rm ab} \left(\frac{\sqrt{s}}{10 \, {\rm TeV}}\right)^{\!2} \!\! \left(\frac{\Delta a_\mu}{3 \times 10^{-9}} \right)^{\!2}
\end{align}
where in the last equality we have again taken $\Lambda \approx 100~$TeV so that 
$|\Delta a_\mu| \approx 3 \times 10^{-9} \left(100 \,{\rm TeV}/\Lambda\right)^2 |C^{\mu t}_T|$.
We estimate the reach on $\Delta a_\mu$ simply assuming an overall 50\% efficiency for reconstructing the top quarks, 
and requiring a statistically significant deviation from the SM $\mu^+\mu^-\to t\bar{t}$ background, which has a cross-section
%
$\sigma_{t\bar{t}}^{\rm SM} \approx 1.7 \,{\rm fb}\,\big(\frac{10\,{\rm TeV}}{\sqrt{s}}\big)^2$.
%
Similarly, if the charm-loop contribution dominates, we can probe 
$|\Delta a_\mu| \approx 3 \times 10^{-9} \left(10 \,{\rm TeV}/\Lambda\right)^2 |C^{\mu c}_T|$ through the process $\mu^+\mu^-\to c\bar c$.
In this case, unitarity constraints on the NP coupling $C_T^{\mu c}$ require a much lower NP scale $\Lambda \lesssim 10$~TeV, 
so that our effective theory analysis will only hold for lower center-of-mass energies.
Combining eq.~(\ref{eq:Delta_a_ell}) and (\ref{eq:mumu_tt}), with $c \leftrightarrow t$, we find that
\begin{align}
\sigma_{c\bar{c}}
\, \approx \, 100 \,{\rm fb} \left(\frac{\sqrt{s}}{3 \, {\rm TeV}}\right)^{\!2} \!\! \left(\frac{\Delta a_\mu}{3 \times 10^{-9}} \right)^{\!2}.
\label{eq:mumu_cc_numerics}
\end{align}
The SM cross-section for $\mu^+\mu^- \!\to c\bar{c}$ 
at $\sqrt{s}= 3~$TeV is $\sim 19$~fb. In figure~\ref{fig:hgamma} we show the 95\% C.L.\ constraints on the top and charm contributions to $\Delta a_\mu$ as red and orange lines, respectively, as a function of the collider energy. Notice that the charm contribution can be probed already at $\sqrt{s} = 1$~TeV, while the top contribution can be probed at $\sqrt{s} = 10$~TeV. 
The simultaneous constraints on the NP couplings $C_{e\gamma}^\mu$ and $C_T^{\mu t}$ are shown in figure~\ref{fig:2D} for a 30~TeV collider.

\medskip

So far, we assumed CP conservation. If however the coefficients $C_{e\gamma}$, $C_{eZ}$ or $C_T$ are complex, 
the muon electric dipole moment (EDM) $d_\mu$ is unavoidably generated.
Since the cross-sections in eq.~\eqref{eq:sigma_hV} and \eqref{eq:mumu_tt} are proportional to the absolute values of the same coefficients, a MC offers a unique opportunity to test also $d_\mu$. The current experimental limit $d_\mu < 1.9 \times 10^{-19}\,e\,$cm was set by the BNL E821 experiment~\cite{Bennett:2008dy} 
and the new E989 experiment at Fermilab aims to decrease this by two orders of magnitude~\cite{Chislett:2016jau}. 
Similar sensitivities could be reached also by the J-PARC $g$-2 experiment~\cite{Gorringe:2015cma}.

From the model-independent relation~\cite{Giudice:2012ms}
\begin{align}
\frac{d_\mu}{\tan\phi_\mu} =  \frac{\Delta a_\mu}{2 m_\mu} \,e \,\simeq\, 3 \times 10^{-22} \left(\frac{\Delta a_\mu}{3\!\times\! 10^{-9}}\right) e\, {\rm cm}\,,
\label{eq:d_mu}
\end{align}
where $\phi_\mu$ is the argument of the dipole amplitude, the bounds on $\Delta a_\mu$ in figure~\ref{fig:hgamma} can be translated into a model-independent constraint on $d_\mu$. We find that already a 10~TeV MC can reach a sensitivity comparable to the ones expected at Fermilab~\cite{Chislett:2016jau} and J-PARC~\cite{Gorringe:2015cma}, while at a 30 TeV collider one gets the bound $d_\mu \lesssim 3\times 10^{-22}\, e$~cm.

\vspace{5pt}

\section{IV. Rare Higgs decays} 
We finally discuss the connection between the lepton $g$-2 and the radiative Higgs decays $h \to \ell^+\ell^-\gamma$. 
Due to the large luminosity, and the growth with energy of the vector-boson-fusion cross-section, a huge number of Higgs bosons 
is expected to be produced at a high-energy lepton collider~\cite{Costantini:2020stv}. In particular, a MC running at $\sqrt{s}= 30~$TeV 
with an integrated luminosity of $90 \, {\rm ab}^{-1}$ will produce $\mathcal{O}(10^8)$ Higgs bosons. 
With the precision of Higgs couplings measurements most likely limited by systematic errors, the main advantage of having such a 
large number of events is the possibility to look for very rare decays of the Higgs.

The dipole operator $O_{e\gamma}$ contributes to 
$h\to \ell^+\ell^-\gamma$ as
\begin{equation}
\!\!\Gamma(h \to \ell^+\ell^-\gamma)_{\mysmall\rm NP}\! = \frac{e m_h^3 m_\ell}{64\pi^3v}\frac{{\rm Re}(C^\ell_{e\gamma})}{\Lambda^2} + \frac{m^5_h}{768\pi^3} \frac{|C^{\ell}_{e\gamma}|^2}{\Lambda^4},\!
\label{eq:radiative_width}
\end{equation}
where the first term comes from the interference with the SM tree-level amplitude.
Combining this expression with eq.~(\ref{eq:Delta_a_ell}) gives $\mathcal{B}(h \to \mu^+\mu^-\gamma)_{\mysmall\rm NP} \approx 5 \times 10^{-10} \left(\frac{\Delta a_\mu}{3\times 10^{-9}} \right)$, and thus
the current muon $g$-2 anomaly cannot be tested at a MC through the process $h \to \mu^+\mu^-\gamma$. 
Instead, $\mathcal{B}(h \to \tau^+\tau^-\gamma)_{\mysmall\rm NP} \approx 10^{-5} \left(\frac{\Delta a_\tau}{5\times 10^{-5}} \right)$, and a sensitivity to 
$\Delta a_\tau$ of order $\Delta a_\tau \lesssim 5\times 10^{-5}$ could be attained by measuring $h \to \tau^+\tau^-\gamma$  with percent precision~\cite{gm2_tau}.

The operator $O_{eZ}$ affects the $h\to \ell^+\ell^- Z$ decay in a way analogous to eq.~\eqref{eq:radiative_width}. 
While the contribution in the $h\to \mu^+\mu^- Z$ channel is still too small to be observed, a measurement of 
$\mathcal{B}(h\to \tau^+\tau^- Z)$ 
at the percent level could be sensitive to values of $\Delta a_\tau \lesssim 10^{-4}$.
It is worth pointing out that at a high-energy lepton collider $\Delta a_\tau$ can also be efficiently probed through the processes 
$\mu^+\mu^- \to \tau^+\tau^-$, and especially $\mu^+\mu^- \to \mu^+\mu^- \tau^+\tau^- (\bar\nu \nu\, \tau^+\tau^-)$ which enjoys 
a very large cross-section driven by vector-boson-fusion~\cite{gm2_tau}. 

\vspace{5pt}

\section{V. Conclusions}  
The muon $g$-2 discrepancy is one of most intriguing hints of new physics emerged so far in particle physics, 
which has recently been reinforced with the confirmation of the BNL result~\cite{Bennett:2006fi} by the E989 experiment at Fermilab~\cite{Abi:2021gix}.
However, these low-energy determinations of $\Delta a_\mu$ rely on the assumption that systematic 
and hadronic uncertainties are under control at the outstanding level of $\Delta a_\mu \sim 10^{-9}$.
Therefore, an independent test of $\Delta a_\mu$, not contaminated by the above sources of uncertainty, is very desirable.

In this work, we have demonstrated that a muon collider running at center-of-mass energies of several TeV can achieve this goal, 
providing a unique, model-independent test of new physics in the muon $g$-2 through the study of the high-energy processes 
$\mu^+\mu^- \to h \gamma,hZ,q\bar{q}$. 
In particular, a 30 TeV collider with the baseline integrated luminosity of 90 ab$^{-1}$ would be able to reach a sensitivity to the 
electromagnetic dipole operator of few$\,\times 10^{-9}$, comparable to the present value of $\Delta a_\mu$. If on the other hand 
the $g$-2 anomaly arises at loop-level from quark-lepton interactions, this could already be tested at a few TeV collider.
Furthermore, we have shown that the current bound on the muon electric dipole moment can be improved by three orders 
of magnitude, down to ${\rm few} \times 10^{-22}\,e\,$cm.

These results rely on measurements with $\mathcal{O}(1)$ accuracy, and thus do not require a precise control of systematic 
or theoretical uncertainties. We stress that our findings are completely model-independent, being formulated in terms of 
the very same effective operators that control the lepton dipole moments. Should the muon $g$-2 anomaly be confirmed by 
forthcoming 
investigations, this would constitute a {\it no-lose} theorem for a multi-TeV muon collider, 
guaranteeing the discovery of new physics directly in high-energy collisions.
Our results add a relevant piece to the already far-reaching potential of a muon collider in high-energy physics.

\medskip

{\it Acknowledgments.} 
We thank M. Passera and A. Wulzer for useful discussions. The work of D.B. was supported in part by MIUR under contract PRIN 2017L5W2PT, and by the INFN grant `FLAVOR'.
P.P. acknowledges partial support by FP10 ITN Elusives (H2020-MSCA-ITN-2015-674896) and Invisibles-Plus (H2020-MSCA-RISE-2015-690575). 

\vspace{3pt}



\begin{thebibliography}{99}

\bibitem{Bennett:2006fi}
  G.~W.~Bennett {\it et al.} [Muon g-2 Collaboration],
  Phys.\ Rev.\ D {\bf 73} (2006) 072003.

\bibitem{Abi:2021gix}
  B.~Abi {\it et al.} [Muon g-2 Collaboration],
  Phys.\ Rev.\ Lett.\  {\bf 126} (2021) no.14,  141801;
%
  T.~Albahri {\it et al.} [Muon g-2 Collaboration],
  Phys.\ Rev.\ A {\bf 103} (2021) no.4,  042208;
%
  T.~Albahri {\it et al.} [Muon g-2 Collaboration],
  Phys.\ Rev.\ D {\bf 103} (2021) no.7,  072002.

\bibitem{Aoyama:2020ynm}
  T.~Aoyama {\it et al.},
  Phys.\ Rept.\  {\bf 887} (2020) 1.

\bibitem{Abe:2019thb}
  M.~Abe {\it et al.},
  PTEP {\bf 2019} (2019) no.5,  053C02.

\bibitem{lattice}
Recently, a lattice QCD collaboration computed the leading hadronic contribution to the muon $g$-2 finding a larger value which weakens the discrepancy 
with the experimental result to $1.6\sigma$~\cite{Borsanyi:2020mff}. However, this increase to the hadronic contribution could lead to tensions with the electroweak 
fit or low-energy $e^+e^- \!\to\! {\rm hadron}$ data~\cite{Passera:2008jk}. Therefore, the results of ref.~\cite{Borsanyi:2020mff} should be confirmed or refuted by 
other lattice QCD studies which are underway. Furthermore, there is also the MUonE experimental proposal at CERN aiming to measure the leading hadronic 
contribution to the muon $g$-2 via $\mu e$ scattering~\cite{Abbiendi:2016xup}.


\bibitem{Borsanyi:2020mff}
S.~Borsanyi, Z.~Fodor, J.~N.~Guenther, C.~Hoelbling, S.~D.~Katz, L.~Lellouch, T.~Lippert, K.~Miura, L.~Parato and K.~K.~Szabo, \textit{et al.}
Nature \textbf{593} (2021) no.7857, 51-55.

\bibitem{Passera:2008jk}
  M.~Passera, W.~J.~Marciano and A.~Sirlin,
  Phys.\ Rev.\ D {\bf 78} (2008) 013009;
%
  A.~Keshavarzi, W.~J.~Marciano, M.~Passera and A.~Sirlin,
  Phys.\ Rev.\ D {\bf 102} (2020) no.3,  033002;
%
  A.~Crivellin, M.~Hoferichter, C.~A.~Manzari and M.~Montull,
  Phys.\ Rev.\ Lett.\  {\bf 125} (2020) no.9,  091801;
%
  G.~Colangelo, M.~Hoferichter and P.~Stoffer,
  JHEP {\bf 1902} (2019) 006;
%
  Phys.\ Lett.\ B {\bf 814} (2021) 136073.

\bibitem{Abbiendi:2016xup}
  G.~Abbiendi {\it et al.},
  Eur.\ Phys.\ J.\ C {\bf 77} (2017) no.3,  139;
%
  C.~M.~Carloni Calame, M.~Passera, L.~Trentadue and G.~Venanzoni,
  Phys.\ Lett.\ B {\bf 746} (2015) 325;
%
  P.~Banerjee {\it et al.},
  Eur.\ Phys.\ J.\ C {\bf 80} (2020) no.6,  591.

\bibitem{Marciano:2016yhf}
  W.~J.~Marciano, A.~Masiero, P.~Paradisi and M.~Passera,
  Phys.\ Rev.\ D {\bf 94} (2016) no.11, 115033.


\bibitem{Buchmuller:1985jz}
  W.~Buchmuller and D.~Wyler,
  Nucl.\ Phys.\ B {\bf 268} (1986) 621;
%
  B.~Grzadkowski, M.~Iskrzynski, M.~Misiak and J.~Rosiek,
  JHEP {\bf 1010} (2010) 085;
%
  E.~E.~Jenkins, A.~V.~Manohar and M.~Trott,
  JHEP {\bf 1401} (2014) 035;
%
  R.~Alonso, E.~E.~Jenkins, A.~V.~Manohar and M.~Trott,
  JHEP {\bf 1404} (2014) 159.

\bibitem{masses}
If the underlying NP generating the dipole moments is also responsible for the generation of the lepton masses, then the chiral enhancement 
for the dipoles is not at work and the involved scales to explain the muon $g-2$ anomaly have to be of the order of a few TeV.

\bibitem{Capdevilla:2020qel}
R.~Capdevilla, D.~Curtin, Y.~Kahn and G.~Krnjaic,
Phys.~Rev.~D~\textbf{103} (2021) no.7, 075028

\bibitem{Delahaye:2019omf}
J.~P.~Delahaye, M.~Diemoz, K.~Long, B.~Mansouli\'e, N.~Pastrone, L.~Rivkin, D.~Schulte, A.~Skrinsky and A.~Wulzer,
[arXiv:1901.06150 [physics.acc-ph]].
  
  


\bibitem{muoncollider}
M.~Bogomilov \textit{et al.} [MICE],
Nature \textbf{578} (2020) no.7793, 53-59;
Bartosik {\it et al.}, JINST {\bf 15} (2020) P05001; 
M. Boscolo, J.~P.~Delahaye and M. Palmer, RAST {\bf 10} (2019) 189;
\href{https://muoncollider.web.cern.ch}{https://muoncollider.web.cern.ch}.  

\bibitem{Costantini:2020stv}
  A.~Costantini {\it et al.}, F.~De Lillo, F.~Maltoni, L.~Mantani, O.~Mattelaer, R.~Ruiz and X.~Zhao,
  JHEP {\bf 2009} (2020) 080;
D.~Buttazzo, D.~Redigolo, F.~Sala and A.~Tesi,
JHEP \textbf{11} (2018), 144;
  S.~Dawson and J.~L.~Rosner,
  Phys.\ Lett.\  {\bf 148B} (1984) 497;
%
%
  G.~Altarelli, B.~Mele and F.~Pitolli,
  Nucl.\ Phys.\ B {\bf 287} (1987) 205;
%
  W.~Kilian, M.~Kramer and P.~M.~Zerwas,
  Phys.\ Lett.\ B {\bf 373} (1996) 135;
%
  J.~F.~Gunion, T.~Han and R.~Sobey,
  Phys.~Lett.~B~{\bf 429}~(1998)~79.

\bibitem{gm2_e}
The electron $g$-2 has been invoked in ref.~\cite{Giudice:2012ms} as a test of the muon $g$-2 anomaly
in NP scenarios where the leptonic $g$-2 do not follow a quadratic scaling with lepton masses.

\bibitem{Giudice:2012ms}
  G.~F.~Giudice, P.~Paradisi and M.~Passera,
  JHEP {\bf 1211} (2012) 113.

\bibitem{Abbasabadi:1995rc}
  A.~Abbasabadi, D.~Bowser-Chao, D.~A.~Dicus and W.~W.~Repko,
  Phys.\ Rev.\ D {\bf 52} (1995) 3919;
%
  A.~Djouadi, V.~Driesen, W.~Hollik and J.~Rosiek,
  Nucl.~Phys.~B~{\bf 491} (1997) 68.

\bibitem{Bennett:2008dy}
  G.~W.~Bennett {\it et al.} [Muon (g-2) Collaboration],
  Phys.\ Rev.\ D {\bf 80} (2009) 052008.
  
\bibitem{Chislett:2016jau}
  R.~Chislett [Muon g-2 Collaboration],
  EPJ Web Conf.\  {\bf 118} (2016) 01005.

\bibitem{Gorringe:2015cma}
  T.~P.~Gorringe and D.~W.~Hertzog,
  Prog.\ Part.\ Nucl.\ Phys.\  {\bf 84} (2015) 73.
  
\bibitem{gm2_tau}
D.~Buttazzo and P.~Paradisi, to appear.  

\end{thebibliography}
\end{document}